\documentclass[aip,apl,reprint,superscriptadress]{revtex4-1}%
\usepackage{graphicx}
\usepackage{bm}
\usepackage{dcolumn}
\usepackage[latin9]{inputenc}
\usepackage{amsmath}

\begin{document}
\title{Monopolelike probes for quantitative magnetic force microscopy: calibration and application}

\author{ S.~ Vock}
 \email{s.vock@ifw-dresden.de}
\author{F. ~Wolny}
\author{ T.~M\"uhl}
\author{R.~Kaltofen}
\author{ L.~Schultz}
\author{ B.~B\"uchner}
\affiliation{IFW Dresden, P.O. Box 270116, D-01171 Dresden, Germany}
\author{C.~ Hassel }
\author{ J.~Lindner}
\affiliation{Physik, Universit\"at Duisburg-Essen, 47048 Duisburg, Germany}
\author{ V.~Neu}
\affiliation{IFW Dresden, P.O. Box 270116, D-01171 Dresden, Germany}

\date{\today}

\begin{abstract}
A local magnetization measurement was performed with a Magnetic Force Microscope (MFM) to determine magnetization in domains of an exchange coupled [Co/Pt]/Co/Ru multilayer with predominant perpendicular anisotropy. The quantitative MFM measurements were conducted with an iron filled carbon nanotube tip, which is shown to behave like a  monopole. As a result we determined an additional in-plane magnetization component of the multilayer, which is explained by estimating the effective permeability of the sample within the $\mu^*$-method. 
\end{abstract}


\maketitle 
The study of ferromagnetic layers with perpendicular anisotropy is of special interest for applications in perpendicular magnetic recording technology. Based on a qualitative study\cite{BRAN09}, a $ [(Co/Pt)_{8}/Co/Ru]_{18}$ multilayer composed of individual blocks of $[Co(0.4~nm)/Pt(0.7~nm)]_{8}Co(0.4~nm)$ separated by thin Ru spacer layers shows strong perpendicular magnetic anisotropy and a ferromagnetic (FM) band domain state in zero-field. This letter focuses on this FM stripe domain state and will give quantitative insight into the magnetization present in the domains themselves.

Magnetic Force Microscopy (MFM) is a powerful tool for qualitative domain imaging on a sub 100 nm nanometer scale. 
Contrast formation in MFM is based on the magnetostatic interaction between the sample stray field and the magnetic tip and is therefore strongly dependent on the tip properties. These properties are not known a priori and even if they are, the calculation of the magnetization pattern from the measured signal is not trivial, since it requires the inverse solution of a 3 dimensional convolution integral. 

In the following it will be shown that the use of iron-filled carbon nanotubes (Fe-CNT) as MFM tips can overcome these obstacles. Such tips posses a monopole type characteristic and therefore allow for simple calibration and subsequent quantitative MFM measurements on samples with unknown perpendicular magnetization. Moreover these tips provide a high magnetic resolution of at least 25 nm \cite{WOLNY10}.

\begin{figure}[tb] \graphicspath{{figures/}}\centering
\includegraphics[scale=1.0, clip]{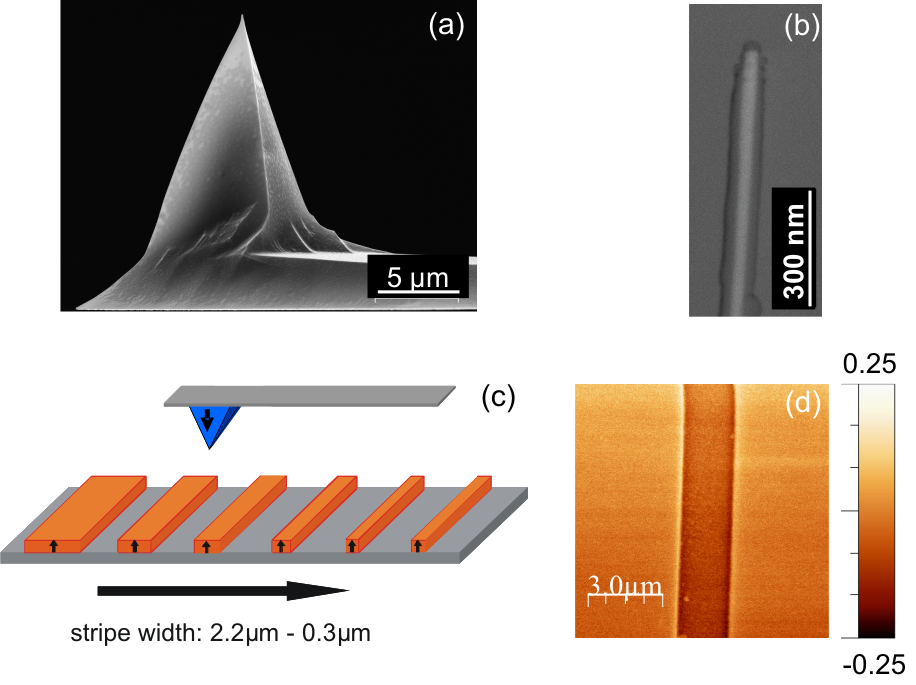}
\caption{SEM image of (a) a MESP type tip (Veeco) and (b) of the Fe-CNT. The sketch in (c) visualizes the calibration sample consisting of 6 CoPt stripes with varying width. (d) MFM image of the 2.2 $\mu m$ stripe.} \label{fig:TipsAndCalibrationSamp}
\end{figure}

The characterization of the MFM tips was performed by calibration measurements on 6 $(Co/Pt)_{7}$ multilayer stripes with varying width from $2.2$ $\mu m$ down to 300 nm and a height of 14 nm (see the sketch in figure \ref{fig:TipsAndCalibrationSamp} (c)). These stripes posses a well defined stray field and are therefore suitable reference samples. The stripes were prepared by alternating electron beam evaporation of Co and Pt  and high resolution electron beam lithography plus lift-off  \cite{HASSEL06}. In-plane magnetoresistance measurements revealed an effective magnetic anisotropy constant of $K_{eff} = 1.05$x$10^6J/m^3$ and a saturation magnetization of $ M_{s}$ = $1080$ $kAm^{-1}$ was obtained by SQUID measurements. With this we find Q values [$Q = K_{eff}/\frac{1}{2}\mu_0 M_S^2$] of about Q = 1.5 which indicates that flux closure domains should not be present. The remnant state after perpendicular saturation is single domain.

Two types of MFM tips were characterized and compared. Fe-CNT tips, which were prepared by chemical vapor deposition (CVD) and attached to a standard AFM cantilever\cite{WOLNY08} and a commercial MESP (trade name: metal coated etched silicon probe from Veeco) type tip. The Fe-CNT consists of a carbon shell with an iron filling of $2$ $\mu m$ in length and a radius of approximately $16$ nm, whereas the MESP tips have a pyramidal shape and are coated with a thin magnetic layer. Both tips are magnetized along the axis perpendicular to the cantilever plane. All MFM measurements were carried out with a Digital Instruments 3100 scanning probe microscope in the tapping/lift mode.

To derive the tip characteristics from the calibration measurement the simple and straightforward point probe approximations \cite{HARTMANN88} were applied. Within these models the probe is described by an effective magnetic charge (q) or moment (m) located in one single point within the probe volume (with the distance $\delta$ from the tip apex) and a mutual influence on the magnetization is excluded. These assumptions allow calculation of the resulting MFM signal $\Delta \Phi$, which is the phase shift of the oscillating cantilever, by a simple multiplication of the probes monopole moment q with the first sample stray field derivative
\begin{equation} 
\Delta \Phi (x,y,z+\delta) = -\mu_0\frac{180^{\circ}}{\pi}\frac{Q}{k}q\frac{\partial H_z}{\partial z}(x,y,z+\delta)
\end{equation}
or the dipole moment \textbf{m} with the second stray field derivative above the sample
\begin{equation}
\begin{split}
\Delta \Phi (x,y,z+\delta)= -\mu_0\frac{180^{\circ}}{\pi}&\frac{Q}{k}\bigg[m_x\frac{\partial^2 H_x}{\partial z^2}(x,y,z+\delta)\\&+m_y\frac{\partial^2
H_y}{\partial z^2}(x,y,z+\delta)\\&+m_z\frac{\partial^2 H_z}{\partial z^2}(x,y,z+\delta)\bigg].
\end{split}
\end{equation}
The quality factor Q can be extracted from resonance characteristics and the spring constant k was calculated out of the cantilever geometry and the young modulus of Si. The stray field derivatives can be calculated for a perpendicularly magnetized stripe from the corresponding topography image. 
With this procedure whole MFM scanlines are modelled and fitted to the measured results. Fitting parameters within the dipole model are the direction and absolute value of the vector \textbf{m} and its position $\delta$ within the tip, whereas in the monopole model only the absolute value of the monopole charge and $\delta$ is optimized.
\begin{figure}[tb] \graphicspath{{figures/}}\centering
\includegraphics[scale=1.0, clip]{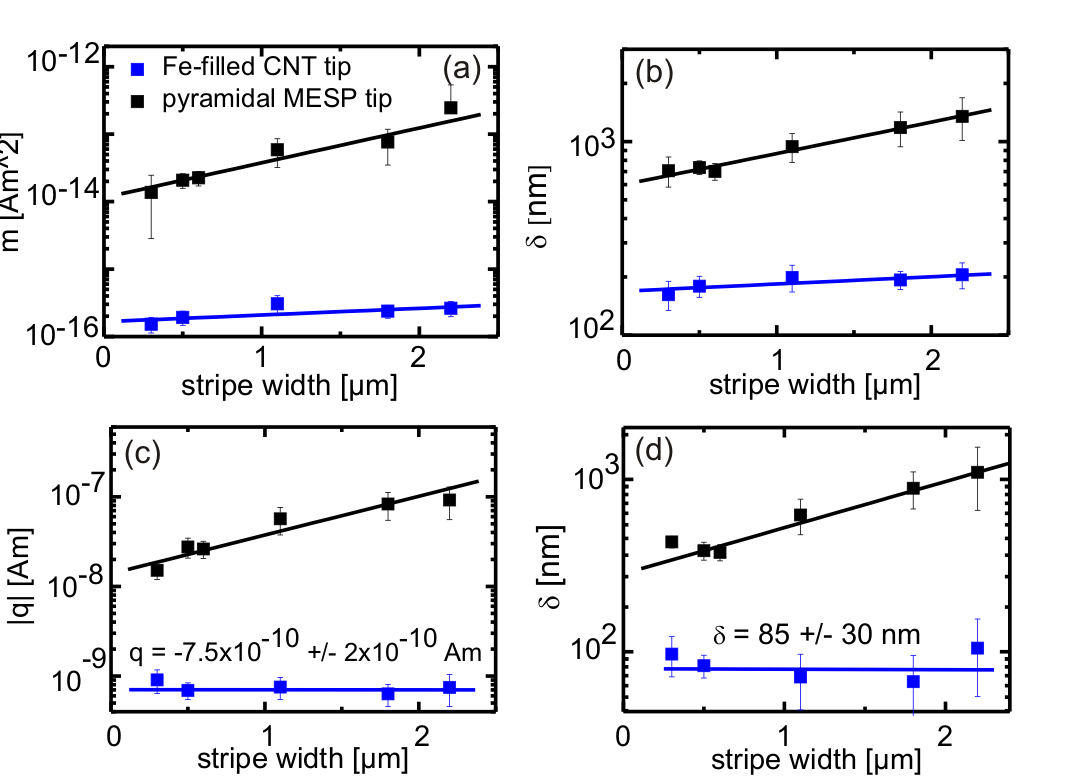}
\caption{Calculated monopole (a) and dipole moments (c) and the corresponding $\delta$ values ((b) and (d)) for the MESP tip and the Fe-filled CNT tip in an average lift height of 60-80 nm. Within the dipole model a tilt of $19^{\circ}$ of the moment in the MESP tip was found and kept constant during the whole calibration procedure.} \label{fig:dipoleandmonopole}
\end{figure}
The MFM signal is averaged over approximately 50 scanlines to improve the signal-to-noise ratio. Before measurement the tips and the stripes were magnetized with a NdFeB bulk magnet in the same direction.

The results of the calibration procedure are summarized in figure \ref{fig:dipoleandmonopole}. Tip calibration parameters m, q and $\delta$ are plotted on a logarithmic scale as a function of the magnetic structure size to accommodate the large span of values resulting from the fit procedure. The MESP type tip shows an increase of the dipole moment and the corresponding $\delta$ with increasing structure size (\ref{fig:dipoleandmonopole} (a) and (b)). The same is true for the monopole model (\ref{fig:dipoleandmonopole} (c) and (d)). The tip parameters of the Fe-CNT tip show a slight increase with increasing stripe width within the dipole model (\ref{fig:dipoleandmonopole} (a) and (b)) and stay constant within the monopole model (\ref{fig:dipoleandmonopole} (c) and (d)). The calibration results are as expected and, for the MESP type tip, in good agreement with values reported earlier for comparable tips \cite{KEBE04,LOHAU99}. They can be understood by means of a changing effective tip volume caused by an increasing decay length of the stray field when increasing the stripe (domain) width. Therefore the magnetic point charge position, which is located in the middle of the effective volume, moves upwards, thus increasing the tip moment. This behaviour holds for all tips which do not resemble a true point pole: the values for moment (m or q) and displacement ($\delta$) necessary to describe the tip response to the sample represent the properties of an extended tip by localizing these into one point. In case of varying stray field geometries the interaction volume in the tip is changing and therefore requires modified tip parameters \cite{LOHAU99}. This makes the application of the point probe approximation to standard MFM tips rather difficult, since the tip has to be calibrated for each stray field configuration separately. In contrast to that, the Fe-CNT tip comes close to a true point monopole, whose properties are not expected to vary for different stray
field geometries. These considerations are confirmed by the calibration data in figure \ref{fig:dipoleandmonopole} (c) and (d). The charge and
$\delta$ are constant for the Fe-CNT tip within the monopole model. Moreover the magnitude of the monopole moment $q$ = ($0.8\pm0.2)$x$10^{-9}$ Am is
comparable to that expected from geometrical considerations ($q_{geom} = M_s^{Fe}\pi r_{tip}^2 = 1.4$x$10^{-9}$ Am). Considering the carbon shell of the nanotube which is in the range of 30 nm a delta of $85\pm30$ nm is reasonable. Small diameter variations at the end of the iron filling can cause the deviation from the cylindrical geometry model. Within the dipole model (figure \ref{fig:dipoleandmonopole} (a) and (b)) m and $\delta$ increase slightly, which results from the magnetostatic definition of the point dipole and the lateral extension of the iron filling (2$\mu m$) in z direction. In contrast to this the iron filling with a radius of only 16 nm constitutes an almost perfect monopole in the x-y plane relative to the stripe dimensions ranging from 300 to 2200 nm. 
From the above findings it can be concluded that for the Fe-CNT tips the monopole approximation is an absolutely adequate model which describes the tip entirely.

\begin{figure}[h] \centering \graphicspath{{figures/}}
\includegraphics[scale = 1.0, clip]{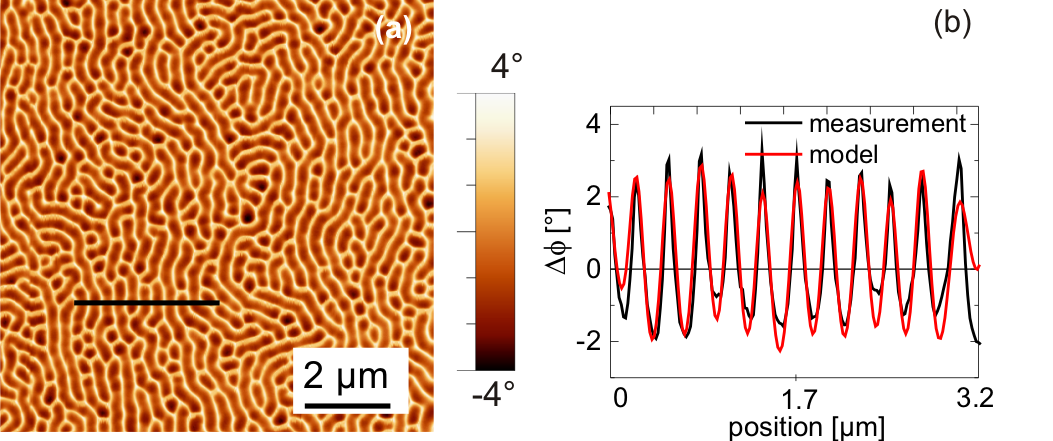}
\caption{(a) MFM image of [Co/Pt]/Co/Ru mulitlayer in the ferromagnetic band domain state. (b) The simulation was performed along the drawn line. The diagram shows the comparison with the simulation.} \label{fig:CoPtRu}
\end{figure}

With the knowledge of the tip properties and its independence on the domain size, we used the Fe-CNT tip for quantitative imaging on a [Co/Pt]/Co/Ru multilayered thin film with the following architecture: $ [(Co(0.4\ nm)/Pt (0.7\ nm))_8/Co(0.4\ nm) /Ru(0.9\ nm)]_{18}$. The film is grown on a 2 nm Pt buffer layer and covered with 2 nm Pt (sputter deposited at $ 1$x$10^{-3}~ mbar$ Ar pressure and the following deposition rates: $Ru = 2.5\ nm/min$, $Pt = 4.4\ nm/min$, $Co = 3.1\ nm/min$). The ferromagnetic Co/Pt multilayer stacks posses a high perpendicular anisotropy. Despite the non-magnetic spacer layer which mediates an antiferromagnetic interlayer exchange coupling between these ferromagnetic layers, in zero field the sample is in ferromagnetic band domain state, where the perpendicular magnetization is correlated in vertical direction throughout the whole block, but forms neighboring domains with opposite magnetization direction and a domain width of about 180 nm (comparable to the sample described in reference 1). The MFM image is shown in figure \ref{fig:CoPtRu} (a).
For the quantification of the measured MFM signal the procedure described previously is applied in the same manner but now keeping the tip properties constant and using the sample magnetization as optimization parameter. Note that in this case we calculated the sample stray field from a charge pattern that was derived by applying a discrimination level to the MFM data. The model assumes zero width domain walls. As tip parameters the mean monopole moment $q_{mean}= -7.54$x$10^{-10}$ Am and the mean displacement $\delta_{mean} = 85 $ nm were applied. The results of the modelling can be found in figure \ref{fig:CoPtRu}. Optimum agreement is found when a sample magnetization $ M = (440 \pm 135) \space kA/m$ is assumed. This value can be compared to global measurements performed with vibrating sample magnetometry which reveal a value of $M_s = 650\pm 65$ $kA/m$.
We obviously resolve an average deviation of each domain from the perfectly perpendicular magnetization orientation expected from the interface anisotropy. This is to a small extent caused by the non-vanishing width of the domain walls. The main contribution, however, comes from the competition between stray field ($K_d$) and anisotropy energy ($K_u$). Only for large perpendicular anisotropy $Q = K_{u}/K_{d} \gg1$ the magnetization in a band domain structure is expected to lie fully perpendicular to the surface within each domain. For smaller $Q\geq1$, the sample adopts a non-homogeneous magnetization structure with considerable in-plane components \cite{STILLRICH10}. In an early model assuming a one-dimensional charge pattern on the surface of an infinite plane Williams et al. \cite{WILLIAMS49} estimated the influence of the effective permeability $\mu^*=1+K_d/K_u$ on the stray field energy and thus the magnetization within each domain. They derived a correction factor $a = 2/(1+\sqrt{\mu^*})$ by which the perpendicular component of the magnetization is expected to change for $\mu^* \neq 1$. The perpendicular uniaxial anisotropy of the $[Co/Pt)/Co/Ru]_{18}$ multilayer was determined by measuring the VSM in-plane (hard) and out-of-plane (easy) hysteresis loops and subtracting the shape anisotropy $K_d = J^2_s/2\mu_0$ from the area enclosed by the hard and easy axis loop \cite{BRANDENBURG09}.
The resulting correction factor is $a = 0.8$, which comes close to the value obtained from the quantitative MFM measurement, where $a = 0.7$.

The results clearly demonstrate the necessity of a complementary measurement technique beside the volume averaging VSM for an overall interpretation of the magnetization status from a microscopic point of view and identify Fe-CNT MFM tips as ideal monopole probes for easy quantitative MFM measurements.

S.V. acknowledges the Cusanuswerk for supporting this work with a scholarship. We would like to acknowledge U. Wolff and C. Bran for helpful discussions and F. Rhein and J.-R. Scholz for help with data analysis.


\begin{thebibliography}{10}%
\makeatletter
\providecommand \@ifxundefined [1]{%
 \ifx #1\undefined \expandafter \@firstoftwo
 \else \expandafter \@secondoftwo
\fi
}%
\providecommand \@ifnum [1]{%
 \ifnum #1\expandafter \@firstoftwo
 \else \expandafter \@secondoftwo
\fi
}%
\providecommand \enquote [1]{``#1''}%
\providecommand \bibnamefont  [1]{#1}%
\providecommand \bibfnamefont [1]{#1}%
\providecommand \citenamefont [1]{#1}%
\providecommand\href[0]{\@sanitize\@href}%
\providecommand\@href[1]{\endgroup\@@startlink{#1}\endgroup\@@href}%
\providecommand\@@href[1]{#1\@@endlink}%
\providecommand \@sanitize [0]{\begingroup\catcode`\&12\catcode`\#12\relax}%
\@ifxundefined \pdfoutput {\@firstoftwo}{%
 \@ifnum{\z@=\pdfoutput}{\@firstoftwo}{\@secondoftwo}%
}{%
 \providecommand\@@startlink[1]{\leavevmode}%
 \providecommand\@@endlink[0]{}%
}{%
 \providecommand\@@startlink[1]{%
  \leavevmode
  \pdfstartlink
   attr{/Border[0 0 1 ]/H/I/C[0 1 1]}%
   user{/Subtype/Link/A<</Type/Action/S/URI/URI(#1)>>}%
  \relax
 }%
 \providecommand\@@endlink[0]{\pdfendlink}%
}%
\providecommand \url  [0]{\begingroup\@sanitize \@url }%
\providecommand \@url [1]{\endgroup\@href {#1}{\urlprefix}}%
\providecommand \urlprefix [0]{URL }%
\providecommand \Eprint[0]{\href }%
\@ifxundefined \urlstyle {%
  \providecommand \doi [1]{doi:\discretionary{}{}{}#1}%
}{%
  \providecommand \doi [0]{doi:\discretionary{}{}{}\begingroup
  \urlstyle{rm}\Url }%
}%
\providecommand \doibase [0]{http://dx.doi.org/}%
\providecommand \Doi[1]{\href{\doibase#1}}%
\providecommand \selectlanguage [0]{\@gobble}%
\providecommand \bibinfo [0]{\@secondoftwo}%
\providecommand \bibfield [0]{\@secondoftwo}%
\providecommand \translation [1]{[#1]}%
\providecommand \BibitemOpen[0]{}%
\providecommand \bibitemStop [0]{}%
\providecommand \bibitemNoStop [0]{.\EOS\space}%
\providecommand \EOS [0]{\spacefactor3000\relax}%
\providecommand \BibitemShut [1]{\csname bibitem#1\endcsname}%
\bibitem{BRAN09}%
  \BibitemOpen
  \bibfield{author}{%
  \bibinfo {author} {\bibfnamefont{C.}~\bibnamefont{Bran}}, \bibinfo {author}
  {\bibfnamefont{A.~B.}\ \bibnamefont{Butenko}}, \bibinfo {author}
  {\bibfnamefont{N.~S.}\ \bibnamefont{Kiselev}}, \bibinfo {author}
  {\bibfnamefont{U.}~\bibnamefont{Wolff}}, \bibinfo {author}
  {\bibfnamefont{L.}~\bibnamefont{Schultz}}, \bibinfo {author}
  {\bibfnamefont{O.}~\bibnamefont{Hellwig}}, \bibinfo {author}
  {\bibfnamefont{U.~K.}\ \bibnamefont{R\"ossler}}, \bibinfo {author}
  {\bibfnamefont{A.~N.}\ \bibnamefont{Bogdanov}},\ and\ \bibinfo {author}
  {\bibfnamefont{V.}~\bibnamefont{Neu}},\ }%
  \bibfield{journal}{%
  \bibinfo {journal} {Phys. Rev. B}\ }%
  \textbf{\bibinfo {volume} {79}},\ \bibinfo {pages} {024430} (\bibinfo {year} {2009})\BibitemShut{NoStop}%
\bibitem{WOLNY10}%
  \BibitemOpen
  \bibfield{author}{%
  \bibinfo {author} {\bibfnamefont{F.}~\bibnamefont{Wolny}}, \bibinfo {author}
  {\bibfnamefont{T.}~\bibnamefont{M\"uhl}}, \bibinfo {author}
  {\bibfnamefont{U.}~\bibnamefont{Weissker}}, \bibinfo {author}
  {\bibfnamefont{K.}~\bibnamefont{Lipert}}, \bibinfo {author}
  {\bibfnamefont{J.}~\bibnamefont{Schumann}}, \bibinfo {author}
  {\bibfnamefont{A.}~\bibnamefont{Leonhardt}},\ and\ \bibinfo {author}
  {\bibfnamefont{B.}~\bibnamefont{B\"uchner}},\ }%
  \bibfield{journal}{%
  \bibinfo {journal} {Nanotechnology}\ }%
  \textbf{\bibinfo {volume} {21}},\ \bibinfo {pages} {435501} (\bibinfo {year}
  {2010})\BibitemShut{NoStop}%
\bibitem{HASSEL06}%
  \BibitemOpen
  \bibfield{author}{%
  \bibinfo {author} {\bibfnamefont{C.}~\bibnamefont{Hassel}}, \bibinfo {author}
  {\bibfnamefont{M.}~\bibnamefont{Brands}}, \bibinfo {author}
  {\bibfnamefont{F.~Y.}\ \bibnamefont{Lo}}, \bibinfo {author}
  {\bibfnamefont{A.~D.}\ \bibnamefont{Wieck}},\ and\ \bibinfo {author}
  {\bibfnamefont{G.}~\bibnamefont{Dumpich}},\ }%
  \bibfield{journal}{%
  \bibinfo {journal} {Phys. Rev. Lett.}\ }%
  \textbf{\bibinfo {volume} {97}},\ \bibinfo {pages} {226805} (\bibinfo {year}
  {2006})\BibitemShut{NoStop}%
\bibitem{WOLNY08}%
  \BibitemOpen
  \bibfield{author}{%
  \bibinfo {author} {\bibfnamefont{F.}~\bibnamefont{Wolny}}, \bibinfo {author}
  {\bibfnamefont{U.}~\bibnamefont{Weissker}}, \bibinfo {author}
  {\bibfnamefont{T.}~\bibnamefont{M\"uhl}}, \bibinfo {author}
  {\bibfnamefont{A.}~\bibnamefont{Leonhardt}}, \bibinfo {author}
  {\bibfnamefont{S.}~\bibnamefont{Menzel}}, \bibinfo {author}
  {\bibfnamefont{A.}~\bibnamefont{Winkler}},\ and\ \bibinfo {author}
  {\bibfnamefont{B.}~\bibnamefont{B\"uchner}},\ }%
  \bibfield{journal}{%
  \bibinfo {journal} {J. Appl. Phys.}\ }%
  \textbf{\bibinfo {volume} {104}},\ \bibinfo {pages} {064908} (\bibinfo {year}
  {2008})\BibitemShut{NoStop}%
\bibitem{HARTMANN88}%
  \BibitemOpen
  \bibfield{author}{%
  \bibinfo {author} {\bibfnamefont{U.}~\bibnamefont{Hartmann}},\ }%
  \bibfield{journal}{%
  \bibinfo {journal} {J. Appl. Phys.}\ }%
  \textbf{\bibinfo {volume} {64}},\ \bibinfo {pages} {1561} (\bibinfo {year}
  {1988})\BibitemShut{NoStop}%
\bibitem{KEBE04}%
  \BibitemOpen
  \bibfield{author}{%
  \bibinfo {author} {\bibfnamefont{T.}~\bibnamefont{Kebe}}\ and\ \bibinfo
  {author} {\bibfnamefont{A.}~\bibnamefont{Carl}},\ }%
  \bibfield{journal}{%
  \bibinfo {journal} {J. Appl. Phys.}\ }%
  \textbf{\bibinfo {volume} {95}},\ \bibinfo {pages} {775} (\bibinfo {year} {2004})\BibitemShut{NoStop}%
\bibitem{LOHAU99}%
  \BibitemOpen
  \bibfield{author}{%
  \bibinfo {author} {\bibfnamefont{J.}~\bibnamefont{Lohau}}, \bibinfo {author}
  {\bibfnamefont{S.}~\bibnamefont{Kirsch}}, \bibinfo {author}
  {\bibfnamefont{A.}~\bibnamefont{Carl}}, \bibinfo {author}
  {\bibfnamefont{G.}~\bibnamefont{Dumpich}},\ and\ \bibinfo {author}
  {\bibfnamefont{E.~F.}\ \bibnamefont{Wassermann}},\ }%
  \bibfield{journal}{%
  \bibinfo {journal} {J. Appl. Phys.}\ }%
  \textbf{\bibinfo {volume} {86}},\ \bibinfo {pages} {3410} (\bibinfo {year}
  {1999})\BibitemShut{NoStop}%
\bibitem{STILLRICH10}%
  \BibitemOpen
  \bibfield{author}{%
  \bibinfo {author} {\bibfnamefont{H.}~\bibnamefont{Stillrich}}, \bibinfo
  {author} {\bibfnamefont{C.}~\bibnamefont{Menk}}, \bibinfo {author}
  {\bibfnamefont{R.}~\bibnamefont{Fr\"omter}},\ and\ \bibinfo {author}
  {\bibfnamefont{H.~P.}\ \bibnamefont{Oepen}},\ }%
  \bibfield{journal}{%
  \bibinfo {journal} {J. Magn. Magn. Mater.}\ }%
  \textbf{\bibinfo {volume} {322}},\ \bibinfo {pages} {1353} (\bibinfo {year}
  {2010})\BibitemShut{NoStop}%
\bibitem{WILLIAMS49}%
  \BibitemOpen
  \bibfield{author}{%
  \bibinfo {author} {\bibfnamefont{H.~J.}\ \bibnamefont{Williams}}, \bibinfo
  {author} {\bibfnamefont{R.~M.}\ \bibnamefont{Bozorth}},\ and\ \bibinfo
  {author} {\bibfnamefont{W.}~\bibnamefont{Shockley}},\ }%
  \bibfield{journal}{%
  \bibinfo {journal} {Phys. Rev.}\ }%
  \textbf{\bibinfo {volume} {75}},\ \bibinfo {pages} {155} (\bibinfo {year}
  {1949})\BibitemShut{NoStop}%
\bibitem{BRANDENBURG09}%
  \BibitemOpen
  \bibfield{author}{%
  \bibinfo {author} {\bibfnamefont{J.}~\bibnamefont{Brandenburg}}, \bibinfo {author}
  {\bibfnamefont{R.}\ \bibnamefont{H\"uhne}}, \bibinfo {author}
  {\bibfnamefont{L.}\ \bibnamefont{Schultz}}, and \bibinfo {author}
  {\bibfnamefont{V.}~\bibnamefont{Neu}},\ }%
  \bibfield{journal}{%
  \bibinfo {journal} {Phys. Rev. B}\ }%
 \textbf{\bibinfo {volume} {79}},\ \bibinfo {pages} {054429} (\bibinfo {year} {2009})\BibitemShut{NoStop}%
\end{thebibliography}
\end{document}